 \documentclass[aps,pra,superscriptaddress,amsmath,amssymb,preprintnumbers,twocolumn,floatfix,showpacs,showkeys,10pt]{revtex4-1}
 \usepackage{amssymb} \usepackage{epsfig}
 \begin{document}
  \title{Relationship between average correlation and quantum steering for arbitrary two-qubit states}

 \author{Youneng Guo}
\affiliation{School of Electronic Communication and Electrical Engineering, Changsha University, Changsha, Hunan
410022, People's Republic of China}
\email{correspond author: guoxuyan2007@163.com}
 \author{Xiangjun Chen}
\affiliation{School of Electronic Communication and Electrical Engineering, Changsha University, Changsha, Hunan
410022, People's Republic of China}
\author{Huping peng}
\affiliation{School of Electronic Communication and Electrical Engineering, Changsha University, Changsha, Hunan
410022, People's Republic of China}
\author{Qinglong Tian}
\affiliation{School of Computer Science and Engineering, Changsha University, Changsha, Hunan
410022, People's Republic of China}

 \begin{abstract}
Quantum nonlocality and nonclassicality are two remarkable characteristics of quantum theory, and offer quantum advantages in some quantum information processing. Motivated by recent work on the interplay between nonclassicality quantified by average correlation [Tschaffon et al., Phys. Rev. Res. 5,023063 (2023)] and Bell nonlocality, in this paper we aim to establish the relationship between the average correlation and the violation of the three-setting linear steering inequality for two-qubit systems. Exact lower and upper bounds of average correlation versus steering are obtained, and the respective states which suffice those bounds are also characterized. For clarity of our presentation, we illustrate these results with examples from well-known classes of two-qubit states. Moreover, the dynamical behavior of these two quantifiers is carefully analyzed under the influence of local unital and nonunital noisy channels. The results suggest that average correlation is closely related to the violation of three-setting linear steering, like its relationship with Bell violation. Particularly, for a given class of states, the hierarchy of nonclassicality-steering-Bell nonlocality is demonstrated.

 \end{abstract}

 \maketitle
\section{Introduction}

Quantum nonlocality and nonclassicality as two remarkable characteristics of quantum theory, are in fact closely related
with each other, especially to capture some nonclassical aspect of quantum states\cite{Jebaratnam}. Nevertheless, there are also subtle differences: The former is more related to outcomes of experimental measurement statistics, and it can actually be viewed as a particular case of nonclassicality, which defies any local hidden variable theory. While the latter has its root in superposition of quantum states encompassing aspects like entanglement(or inseparability)\cite{Horodecki} and squeezing\cite{Zhong}, providing a quantum advantage for computing and information processing tasks compared to classical algorithm, such as quantum key distribution\cite{Phys.Rev.Research6, Phys.Rev.Applied20}, entanglement-based quantum cryptography\cite{Nature582(2020)}, boson sampling\cite{Phys.Rev.Lett.127(2021)}, and so on.

The similarities and differences between these two concepts have attracted much attention from many different perspectives. One of the most fascinating research topics is to characterize and quantify quantum nonlocality and nonclassicality of quantum states. Consequently, various measures have been proposed in the past few years\cite{Phys.Rev.A44(1991),Phys.Rev.A86(2012),Phys.Rev.Lett.80(1998),Rev.Mod.Phys.92(2020),Phys.Rev.Lett.88(2001),Phys.Rev.A77(2008),Rev.Mod.Phys.86(2014),Phys.Rev.Lett.94(2005),Phys.Rev.Lett.98(2007)}. For example, Bell inequality\cite{Bell} often serves as a sufficient but not necessary condition for nonlocality( e.g., the Clauser-Horne-Shimony-Holt inequality\cite{Clauser}): $|\langle\mathfrak{B}_{CHSH}\rangle_{\rho}|\leq2$, where $\mathfrak{B}_{CHSH}=\vec{a}\cdot\vec{\sigma}\otimes(\vec{b}+\vec{b'})\cdot\vec{\sigma}+\vec{a'}\cdot\vec{\sigma}\otimes(\vec{b}-\vec{b'})\cdot\vec{\sigma}$ depends on four measurement directions $(\vec{a}, \vec{a'}, \vec{b} , \vec{b'})$. This inequality holds true for all putative local hidden variable theories, but can be violated for some nonlocal states in quantum mechanics. However, the situation is not always true for some nonclassical states that may obey this inequality\cite{Rev.Mod.Phys.86(2014)}. To better understand quantum nonlocality and nonclassicality of quantum states, one of the most popular nonclassical measures named quantum discord\cite{Phys.Rev.Lett.88(2001)} which quantifies all nonclassical correlations between subsystems in a quantum system is developed.

More recently, Tschaffon et al.\cite{Phys.Rev.Res.5(2023),Phys.Rev.Res.6(2024)} have proposed a necessary and sufficient condition for nonclassicality in terms of average correlation. Specifically, a necessary condition for nonclassicality is required $\Sigma(\rho)\geq\frac{1}{4}$, whereas a sufficient condition for nonclassicality is $\Sigma(\rho)\geq\frac{1}{2\sqrt{2}}$. Even though the above mentioned Bell inequality and quantum discord are used to detecting nonclassical correlations, it's worth emphasizing here that this concept of average correlation differs from Bell inequality or other nonclassicality measures. For example, the violation of Bell inequality is deeply depended on the measurement directions and the underlying quantum state, while other measures such as quantum discord suffer from an extremum over all measurements. In contrast, average correlation based on randomized measurements has the operational and calculable advantages that do not require good control of measurement directions.

Beyond the characterization and quantification of nonlocality and nonclassicality, another interesting research topics is to address the relationship between them. To begin with, an interesting finding in Ref.\cite{Phys.Lett.A(1995)} is that violating a Bell inequality (nonlocality) implies some amount of nonclassicality quantified by concurrence $C$. Subsequently, an exact bound between Bell violation and concurrence for states with a given concurrence is bounded by the relation\cite{Phys.Rev.Lett.89(2002)}: $ 2\sqrt{2}C\leq \mathfrak{B}_{CHSH} \leq 2\sqrt{1+C^2}$. Analogous to the relation between Bell nonlocality and concurrence, the more general relations between Bell nonlocality and several entanglement witnesses according to the negativity, and the relative entropy of entanglement have also been explored\cite{Phys.Rev.A72(2005),Phys.Rev.A81(2010),Phys.Rev.a87(2013),Phys.Rev.A107(2023)}. In Refs.\cite{JPAM44(2011),EPJD66(2012)} where the relationship between the violation of Bell inequality and geometric measure of quantum discord $D_{G}$ for Bell diagonal states have also been derived by the relation: $ 4\sqrt{D_{G}}\leq \mathfrak{B}_{CHSH} \leq 2\sqrt{1+2D_{G}}$. In addition, developments along this line, nonlocality-nonclassicality relation was generalized to the coherence case where the quantitative relation between the Bell nonlocality and the first-order coherence is established by an equality for the pure-state\cite{Nat.7(2012)}, but by an inequality for the mixed-state\cite{Phys.Rev.Lett.115(2015)}.
Very recently, the connection of quantum nonlocality quantified by the maximal violation of the Bell's inequality to average correlation has been studied~\cite{Phys.Rev.Res.5(2023)}. The result shows that violation of Bell's inequality $|\langle\mathfrak{B}_{CHSH}\rangle_{\rho}|>2$ implies the average correlation $\Sigma\geq\frac{1}{4}$, but the presence of average correlation $\Sigma\geq\frac{1}{4}$ does not necessarily imply violation of the inequality.

Inspired by the above relations of nonlocality to nonclassicality and to coherence, this
will permit us to address the relationship between average correlation and another nolocality, quantified by quantum steering.
Quantum steering\cite{Schrodinger}, first proposed by Schr$\ddot{o}$dinger in 1935 to discuss the EPR paradox, is one kinds of nonlocal correlations, lying between Bell nonlocality and entanglement according to the hierarchy of nonlocal correlations. Interest in the relevance of quantum steering ideas to its interplay has increased in recent times\cite{Phys.Rev.A102(2020),Phys.Rev.A103(2021),Phys18(2023)}. One of motivations of this work is to find some similar but alternative relations between average correlation and the maximum violation of the three-setting linear steering
inequality. For a given value of the three-setting linear steering, the analytical expression of average correlation is derived. Specifically, the exact lower and upper bounds of average correlation versus the three-setting linear steering are obtained for a two-qubit system, and the respective states which have extremal bounds are also characterized. To further insight into their connections, we demonstrate the dynamical behavior of these two quantifiers in the presence of noisy channels, including local unital and nonunital channels. Our results show that average correlation is closely related to three-setting linear steering, like its relationship with Bell violation. In particular, for a given class of states, the hierarchy of nonclassicality-steering-Bell nonlocality is demonstrated.

This manuscript is organized as follows. In Sec.II, we give a brief introduction of average correlation and quantum steering. For arbitrary two-qubit state, the link of average correlation to the three-setting linear steering is investigated in Sec.III. In particular, the exact lower and upper bounds of average correlation versus the three-setting linear steering are obtained. For illustration, we consider three examples and compute the analytical expression of average correlation for a fixed amount of steering
 in Section IV.  We compare the dynamical behaviors of the average correlation with that of the quantum steering under decoherence environment. Section V is the conclusion and discussion.

\section{Preliminaries}

Before revealing the relationship between average correlation and quantum steering, we first review their concepts and definitions and present some of the recent results. Without loss of generality, we restrict our attention to a general two-qubit states which can be expressed in the Hilbert-Schmidt decomposition as
\begin{eqnarray}\label{eq}
  \rho_{AB} & = & \frac{1}{4}\left(\openone\otimes\openone+\mathbf{r}\cdot\mathbf{\sigma}\otimes\openone+\openone\otimes \mathbf{s}\cdot\mathbf{\sigma}+\sum_{i,j=1}^{3}t_{ij}\sigma_{i}\otimes\sigma_{j}\right)
\end{eqnarray}
where the Bloch vectors $\mathbf{r}$ and $\mathbf{s}$ belong to $R^{3}$, element $t_{ij}=Tr[\rho(\sigma_{i}\otimes\sigma_{j})]$ form a $3\times 3$ real matrix denoted by $T$ presents the correlation matrix. According to the results obtained in Ref.\cite{Phys.Rev.A62(2020)}, one can find a product unitary transformation $U\otimes V$ which can transform $\rho_{AB}$ to $\varrho_{AB}$ with a diagonal correlation matrix $T'$:
\begin{eqnarray}\label{eqe0}
  \varrho_{AB} & = & (U\otimes V)\rho_{AB}(U^{\dagger}\otimes V^{\dagger})\nonumber \\
   & = &\frac{1}{4}\left(\openone\otimes\openone+\mathbf{r'}\cdot\mathbf{\sigma}\otimes\openone+\openone\otimes \mathbf{s'}\cdot\mathbf{\sigma}+\sum_{i=1}^{3}c_{i}\sigma_{i}\otimes\sigma_{i}\right)
\end{eqnarray}
Here $r'=Tr[U(\mathbf{r}\cdot\mathbf{\sigma})U^{\dagger}\sigma]$ and $s'=Tr[V(\mathbf{s}\cdot\mathbf{\sigma})V^{\dagger}\sigma]$ are
the transformed local Bloch vectors.

\subsection{Average Correlation}

Recently, Tschaffon et al.~\cite{Phys.Rev.Res.5(2023)} have introduced an alternative nonclassical quantifier named as average correlation which is defined as the average absolute value of the two-qubit correlation function
\begin{equation}\label{EQ1}
    \Sigma(\rho)=\frac{1}{4\pi^2}\int d\Omega_\textbf{a} \int d\Omega_\textbf{b} |E(\textbf{a},\textbf{b})|
\end{equation}
where $E(\textbf{a},\textbf{b})=Tr[\rho_{AB}\textbf{a}\cdot\mathbf{\sigma}\otimes \textbf{b}\cdot\mathbf{\sigma}]$ denotes the correlation function measured by two observers averaging over all measurement directions \textbf{a} and \textbf{b}. To evaluate this quantity, one can rewrite the correlation function as
\begin{equation}\label{eq2}
    E(a,b)=\textbf{a}^T T \textbf{b}
\end{equation}
where the correlation matrix $T=Tr[ \rho_{AB}\mathbf{\sigma}\otimes\mathbf{\sigma}]$ always exists a decomposition with nonnegative singular values~\cite{Phys.Rev.A104(2021)}: $diag(\alpha,\beta,\gamma)=UT'V^{T}$. By direct comparison of Eq.~(\ref{eqe0}) with  Eq.~(\ref{eq2}), we infer that these nonnegative singular values are identical to $|c_{i}|$ with the order $0\leq\gamma\leq\beta\leq\alpha\leq1$.
Substituting Eq.~(\ref{eq2}) into Eq.~(\ref{EQ1}), after a straightforward calculation, one can obtain the average correlation
\begin{equation}\label{eq3}
\Sigma(\rho) = \frac{\alpha}{8\pi}\int_0^{2\pi}d\phi\int_0^{\pi}d\theta \sin\theta\sqrt{f(\phi)\sin^2\theta+\cos^2\theta}
\end{equation}
with the function
\begin{equation}\label{eq4}
f(\phi)=\left(\frac{\beta}{\alpha}\right)^2\sin^2 \phi+\left(\frac{\gamma}{\alpha}\right)^2\cos^2 \phi.
\end{equation}
As shown in Ref.\cite{Phys.Rev.Res.5(2023)} the average correlation is regarded as a good indicator for nonclassicality ranging from a minimum of $\Sigma=1/4$ to a maximum of $\Sigma =1/2$. Unlike other nonclassicality measures, such as quantum discord and measurement-induced disturbance which involve maximal or minimal procedure over all measurements, average correlation has the operational advantage that does not require precise measurement due to the fact that its definition is averaging absolute value of measuring correlations at random. Therefore, average correlation can be seen as a quantify independent of any shard reference frame.

\subsection{Quantum steering\label{sec:GMQD}}

Quantum steering\cite{Schrodinger} first introduced by Schr$\ddot{o}$dinger in 1935, which captures the ability of an observer on one side by the local measurements to remotely steer the state on the other side, is today recognized as a kind of nonlocal correlation lying between Bell nonlocality and entanglement. Steerable states can be certified by the violation of steering inequality\cite{Phys.Rev.A80(2009),Phys.Rev.A87(2013),JOSA(2015),ScientificReports(2021)} and have been shown to offer quantum advantages in quantum information tasks.

A linear steering inequality based on a finite sum of bilinear expectation values was introduced by Cavalcanti et al. to diagnose the steerability of a quantum state\cite{Phys.Rev.A93(2016)}. For a given two-qubit state, steering inequality can be expressed as
\begin{eqnarray}\label{eq5}
F_{n}(\varrho,\mu)=\frac{1}{\sqrt{n}}\left|\sum_{i=1}^{n}\langle A_{i}\otimes B_{i}\rangle\right|\leq 1,
\end{eqnarray}
where $A_{i}=\mathbf{a}_{i}\cdot \mathbf{\sigma}$ and $B_{i}=\mathbf{b}_{i}\cdot \mathbf{\sigma}$, are Hermitian
operators acting on qubits $A$ and $B$, respectively. Both $\mathbf{a}_{i}$ and $\mathbf{b}_{i}\in R^{3}$ are two unit vectors, and $\mu=\{\mathbf{a}_{i},\mathbf{b}_{i}\}$ is the set of measurement directions. Violation of this inequality implies that the state $\varrho$ is steerable and the quantification of steering is given
by the maximum violation
\begin{eqnarray}\label{eq6}
\mathbb{S}_{n}(\varrho)=\left\{0,\frac{s_{n}-1}{\sqrt{n}-1}\right\},
\end{eqnarray}
Note that quantum steering is fundamentally asymmetrical and $\mathbb{S}_{n}(\varrho)\in[0,1] $ is normalized by introducing a factor $\sqrt{n}-1$. For any two-qubit states in the
Hilbert-Schmidt, the degree of steerability is
\begin{eqnarray}\label{eq7a}
s_{n}=\left\{\begin{array}{cc}
\sqrt{\alpha^2+\beta^2}, & \mbox{ for } \, n=2, \\
\sqrt{\alpha^2+\beta^2+\gamma^2}, & \mbox{ for } \, n=3.
\end{array}\right.
\end{eqnarray}
It is easy to conclude that $s_{3}$ is always larger than $s_{2}$ for any $\varrho$, this means that there exists the hierarchy of witnesses of nonlocality. In other words, the presence of $s_{2}$ to witness nonlocality guarantees the presence of $s_{3}$ for steering. However, it is worth noting that there is not a clear quantitative relationship between $\mathbb{S}_{2}$ and
$\mathbb{S}_{3}$. Here $n=2,3$ is corresponding to the two or three-setting measurements per site.

\section{Extremal average correlation for a fixed quantum steering}
By comparing the notions of average correlation with quantum steering defined in Eqs.~(\ref{eq3}) and ~(\ref{eq6}), both of which are put
forward to characterize some kind of nonclassical correlations from different perspectives, this allows us to establish the possible relation between them. Substituting Eq.~(\ref{eq7a}) into  Eq.~(\ref{eq4}), we have
\begin{eqnarray}
f(\phi)=\left\{\begin{array}{cccc}
\frac{s_{3}^{2}-\alpha^2-\gamma^2}{\alpha^2}\sin^2 \phi+\frac{\gamma^2}{\alpha^2}\cos^2 \phi, & \mbox{ for } \, n=3,  \nonumber \\
\frac{s_{2}^{2}-\alpha^2}{\alpha^2}\sin^2 \phi+\frac{\gamma^2}{\alpha^2}\cos^2 \phi, & \mbox{ for } \, n=2 \nonumber
\end{array}\right.
\end{eqnarray}
leading to the general expression of average correlation for any tow-qubit states with a fixed quantum steering
\begin{equation}\label{eqA1}
\Sigma(\rho) = \frac{\alpha}{4}\left[1+\frac{1}{2\pi}\int_0^{2\pi}d\phi\frac{f(\phi)}{\sqrt{1-f(\phi)}}Arcsinh\left(\sqrt{\frac{1-f(\phi)}{f(\phi)}}\right)\right]
\end{equation}
which only depends on parameters $\alpha, \beta, \gamma$. This allows us to obtain the bound of average correlation over $\alpha, \beta, \gamma$ for a fixed quantum steering $s_{n}$.
To present the extremal average correlation, we first consider the three-setting linear quantum steering($s_{3}=\sqrt{\alpha^2+\beta^2+\gamma^2}$). In this case,
we find the function $f(\phi)$ is monotonically increasing in $\gamma$ since
\begin{equation}\label{eqA02}
\frac{\partial f}{\partial\gamma}=\frac{\gamma[s_{3}^{2}+(s_{3}^{2}-2\beta^2)\cos2\phi]}{(s_{3}^{2}-\beta^2-\gamma^2)^2}\geq0
\end{equation}
holds for any $\gamma, \beta, \phi$. Indeed, Eq.~(\ref{eqA02}) can be easily confirmed from $0\leq s_{3}^{2}-2\beta^2\leq s_{3}^{2}$. On the other hands, the integrand function $\frac{f(\phi)}{\sqrt{1-f(\phi)}}Arcsinh\left(\sqrt{\frac{1-f(\phi)}{f(\phi)}}\right)$ is also
monotonically increasing in $f(\phi)$. Consequently, the composite function $\Sigma(\rho)$ with respect to $\gamma$ is also monotonically increasing. In this case, the average correlation $\Sigma(\rho)$ becomes minimal at $\gamma=0$ and maximal at $\gamma=\beta$. (Note that there exists the relation $0\leq\gamma\leq\beta\leq\alpha\leq1$.)

When $\gamma=\beta$, the function $f(\phi)$ reduces to $\frac{s_{3}^{2}-\alpha^2}{2\alpha^2}$ which is independent of $\phi$. As a result,
Eq.~(\ref{eqA1}) reduces to
\begin{equation}\label{eq002}
\Sigma(\rho) = \frac{s_{3}}{4\sqrt{2f+1}}\left[1+\frac{f}{\sqrt{1-f}}Arcsinh\left(\sqrt{\frac{1-f}{f}}\right)\right]
\end{equation}
as a function of $f$ and $s_{3}$.  To determine $\Sigma_{max}(\rho)$, we find Eq.~(\ref{eq002}) is monotonically increasing in $f$ due to the fact that $\frac{d}{df}\Sigma(\rho)\geq 0$ for all $f\in[0,1]$. Hence, we then conclude that $\Sigma(\rho)$ has a maximal value for $f=1$, namely,
\begin{equation}
\lim_{f\rightarrow1}\frac{1}{\sqrt{2f+1}}\left[1+\frac{f}{\sqrt{1-f}}Arcsinh\left(\sqrt{\frac{1-f}{f}}\right)\right]=\frac{2}{\sqrt{3}}
\end{equation}
leads to the upper bound of average correlation $\Sigma(\rho)$ for a fixed $s_{3}$ is
\begin{equation}
\Sigma_{max}(\rho) = \frac{s_{3}}{2\sqrt{3}}
\end{equation}
which is strictly monotonically increasing in $s_{3}$.

Let's come back to Eq.(9) where $s_{3}^{2}=\alpha^2+2\beta^2=\alpha^2+2\gamma^2$, and $f=\frac{s_{3}^{2}-\alpha^2}{2\alpha^2}=1$, the states which suffice the upper bound of average correlation are obtained with $\alpha=\beta=\gamma=\frac{s_{3}}{\sqrt{3}}$, here $0\leq s_{3} \leq\sqrt{3}$.

Next we proceed to the minimization of average correlation for a fixed steering parameter $s_{3}$. As we have discussed above, there exists the minimal $\Sigma(\rho)$ at $\gamma=0$. Based on the result obtain Ref.\cite{Maas}, Eq.~(\ref{eqA1}) is equivalent to
\begin{eqnarray}
\Sigma(\rho) = \frac{\alpha}{4}\int_{0}^{\pi/2}\sqrt{1-\left(2-\frac{s_{3}^{2}}{\alpha^2}\right)\sin^2\phi}d\phi
\end{eqnarray}
which only depends on $\alpha$ when fixed $s_{3}$. Thanks to the monotonic decreasing function $\Sigma(\rho)$ respect to $\alpha$, namely $\frac{d }{d\alpha}\Sigma(\rho)\leq0$, there exists a minimal value $\Sigma(\rho)$ located at $\alpha=1$ for $s_{3}\geq 1$ or $\alpha=s_{3}$ for $s_{3}<1$.  Come back to
Eq.(15), we can derive the lower bound of average correlation $\Sigma(\rho)$ for a fixed $s_{3}$
\begin{eqnarray}\label{eq8}
\Sigma_{min}(\rho)=\left\{\begin{array}{cc}
\frac{s_{3}}{4}, & \mbox{ if } \, s_{3}<1  \\
\frac{1}{4}E(s_{3},\phi), & \mbox{ if } \,s_{3}\geq 1
\end{array}\right.
\label{eq0}
\end{eqnarray}
where $E(s_{n},\phi)=\int_{0}^{\pi/2}\sqrt{1-(2-s_{n}^{2})\sin^2\phi}d\phi$. Meanwhile, the lower bound of average correlation is saturated when the states
 with $\beta=\gamma=0$, $\alpha=s_{3}$ for $s_{3}<1$ and $\alpha=1$, $\beta=\sqrt{s_{3}^{2}-1}$, $\gamma=0$ for $s_{3}\geq1$ are satisfied.

So far we have presented analytical results involving the maximum violation of the three-setting linear steering inequality versus the average correlation as illustrated in Fig. 1. Some remarks are referred to the maximum violation of the two-setting linear steering inequality where $s_{2}=\sqrt{\alpha^2+\beta^2}$. Similarly to the before analyze of the relation between average correlation $\Sigma(\rho)$ and $s_{3}(\rho)$, the exact lower and
upper bounds of the average correlation $\Sigma(\rho)$ versus the two-setting linear steering $s_{2}(\rho)$ are bounded by the relation $Min\{\frac{s_{2}}{4},\frac{E(s_{2},\phi)}{4}\}\leq \Sigma(\rho) \leq \frac{s_{2}}{2\sqrt{2}}$, which is in agreement with
the result presented in Ref.\cite{Phys.Rev.Res.5(2023)}. Indeed, the maximal violation of the two-setting linear steering
inequality is equal to the maximum violation of the Bell inequality\cite{Phys.Rev.A93(2016)}, namely $\mathbb{S}_{2}(\varrho)=\mathbb{N}_{2}(\varrho)$.
This allows us to establish the hierarchy of $\mathbb{S}_{2}(\varrho)=\mathbb{N}_{2}(\varrho)\Rightarrow \mathbb{S}_{3}(\varrho) \Rightarrow\Sigma(\varrho)$, meaning that the existence of Bell nonlocality implies steering, which in turn implies nonclassicality, while the converse is not true.

In conclusion, the general relation between the average correlation and quantum steering
is bounded by $Min\{\frac{s_{n}}{4},\frac{E(s_{n},\phi)}{4}\}\leq \Sigma(\rho) \leq \frac{s_{n}}{2\sqrt{n}}$,
this being the main contribution of this paper.

\begin{figure}[htbp]\centering
\includegraphics[width=8cm]{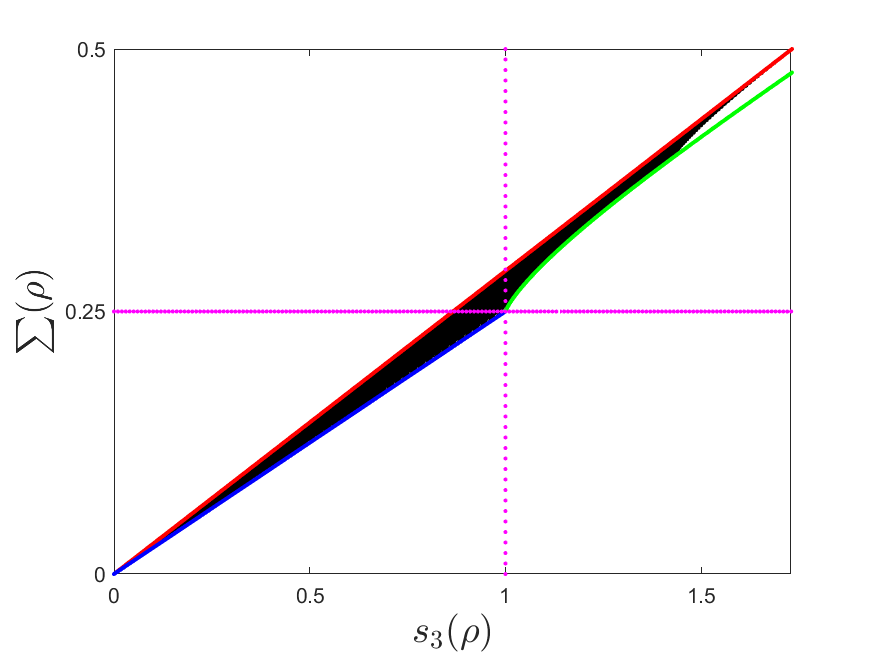}\quad
 \caption{\label{figure1}(Color online) Average correlation $\Sigma(\rho)$ as a function of three-setting linear quantum steering $s_{3}(\rho)$ for two fundamental boundaries: The upper boundary (Red line) represents the function $\Sigma(\rho)=\frac{s_{3}}{2\sqrt{3}}$ and the lower boundary
depicts the curves $\Sigma(\rho)=\frac{s_{3}}{4}$ for $s_{3}<1$(Blue line) and $\Sigma(\rho)=\frac{E_{\phi}}{4}$ for $s_{3}\geq1$(Green line). Obviously, quantum nonlocality quantified by $s_{3}(\rho)\geq1$(equivalent to $\mathbb{S}_{3}(\rho)\geq 0$) implies quantum steering, which in turn implies nonclassicality quantified by $\Sigma(\rho)\geq \frac{1}{4}$. We numerically checked over $10^{5}$ randomly generated bipartite qubit states which all are located between these two boundaries(Black dot).}
\end{figure}

\section{Example}
For illustration, we following will consider three examples where we apply the results mentioned above.
\subsection{Entangled pure states}
Starting with entangled pure states in Schmidt decomposition:
\begin{subequations}
\begin{eqnarray}
|\Psi_{\pm}\rangle =c|0\rangle_A |1\rangle_B \pm \sqrt{1-c^2}|1\rangle_A |0\rangle_B  \\
|\Phi_{\pm}\rangle =c|0\rangle_A |0\rangle_B \pm \sqrt{1-c^2}|1\rangle_A |1\rangle_B,
\end{eqnarray}
\end{subequations}
for which the corresponding correlation matrixes are $diag(\pm2c\sqrt{1-c^2},\pm2c\sqrt{1-c^2},-1)$ and $diag(\mp2c\sqrt{1-c^2},\mp2c\sqrt{1-c^2},1)$, respectively. Obviously, the corresponding to singular values are $\alpha=1$ and $\beta=\gamma=2c\sqrt{1-c^2}$ leading to $s_{3}=\sqrt{1+8c^2(1-c^2)}$. On the other hand, according to Eq.(5), the function $f(\phi)=\frac{1}{2}(s_{3}^{2}-1)$, we can arrive at average correlation
\begin{equation}
\Sigma(\rho) = \frac{1}{4}\left[1+\frac{\sqrt{2}(s_{3}^{2}-1)}{2\sqrt{3-s_{3}^{2}}}Arcsinh\left(\sqrt{\frac{3-s_{3}^{2}}{s_{3}^{2}-1}}\right)\right]
\end{equation}
which is a monotonically increasing respect to $s_{3}$ for pure entangled states.

\subsection{Werner states}
Next, we discuss Werner states\cite{Phys.Rev.A40(1989)}, a probabilistic mixture of maximally mixed separable
state and maximally entangled pure state
\begin{equation}
    \rho=\lambda|\psi_{\pm}\rangle\langle\psi_{\pm}| +\frac{1-\lambda}{4}\openone
\end{equation}
with $|\psi_{\pm}\rangle =\frac{1}{\sqrt{2}}(|0\rangle_A |1\rangle_B \pm |1\rangle_A |0\rangle_B)$. Clearly, the corresponding correlation matrix is $T=-\lambda\openone$ leading to three equal singular values $\alpha=\beta=\gamma=\lambda$. In this case, we specify the function $f(\phi)=1$ and $s_{3}=\sqrt{3}\lambda$, the relationship between average correlation and $s_{3}$ is derived as
\begin{equation}
\Sigma(\rho) = \frac{s_{3}}{2\sqrt{3}}
\end{equation}
It is interesting to note that the upper bound of average correlation is satisfied by Werner states.

\subsection{Maximally entangled mixed states}
The last example is maximally entangled mixed state which is defined as\cite{Phys.Rev.A64(2001)}
\begin{subequations}
\begin{eqnarray}\rho_{I}(s)= \left(
\begin{array}{ c c c c l r }
1/3 & 0 & 0 & s/2 \\
0 & 1/3 & 0 & 0 \\
0 & 0 & 0 & 0 \\
s/2 & 0 & 0 & 1/3 \\
\end{array}
\right) \mbox{ for } 0\leq s \leq \frac{2}{3},  \\
\rho_{II}(s)= \left(
\begin{array}{ c c c c l r }
s/2 & 0 & 0 & s/2 \\
0 & 1-s & 0 & 0 \\
0 & 0 & 0 & 0 \\
s/2 & 0 & 0 & s/2  \\
\end{array}
\right) \mbox{ for } \frac{2}{3}< s \leq 1.
\end{eqnarray}
\end{subequations}
Obviously, for $\rho_{I}(s)$ with $0\leq s \leq \frac{2}{3}$, it is straightforward to
obtain correlation matrix $T=diag(s,s,1/3)$ and $s_{3}=\sqrt{2s^2+1/9}$. For $0\leq s < \frac{1}{3}$, where $\alpha=1/3$ and $\beta=\gamma=s$. The corresponding average correlation reads
\begin{eqnarray}
\Sigma(\rho_{I})=
\frac{1}{12}\left[1+\frac{9s^2}{\sqrt{1-9s^{2}}}Arccsch\left(\frac{3s}{\sqrt{1-9s^2}}\right)\right].
\end{eqnarray}
For $\frac{1}{3}\leq s \leq \frac{2}{3}$, where $\alpha=\beta=s$ and $\gamma=1/3$ lead to
\begin{eqnarray}
\Sigma(\rho_{I})=
\frac{s}{4}\left[1+\frac{1}{2\pi}\int_0^{2\pi}d\phi\frac{f(\phi)}{\sqrt{1-f(\phi)}}Arcsinh\left(\sqrt{\frac{1-f(\phi)}{f(\phi)}}\right)\right]\nonumber \\
\end{eqnarray}
with $f(\phi)=\sin^2 \phi+\frac{1}{9s^2}\cos^2 \phi$.

However, for $\rho_{II}(s)$ with $\frac{2}{3}< s \leq 1$, we find the corresponding correlation matrix $T=diag(s,s,2s-1)$ and $s_{3}=\sqrt{6s^2-4s+1}$. In this case, $\alpha=\beta=s$ and $\gamma=2s-1$ lead to the
average correlation
\begin{eqnarray}
\Sigma(\rho_{II})=\Sigma(\rho_{I})
\end{eqnarray}
with $f(\phi)=\sin^2 \phi+\frac{(2s-1)^2}{s^2}\cos^2 \phi$.
\begin{figure}[htbp]\centering
\includegraphics[width=4cm]{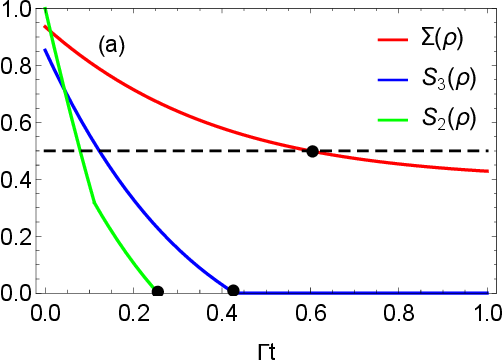}\quad
\includegraphics[width=4cm]{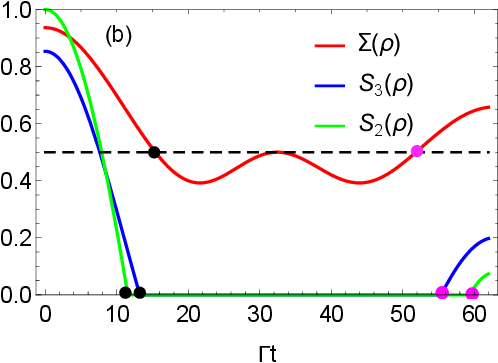}\quad
 \caption{\label{figure2}(Color online)Normalized average correlation  $2\Sigma(\varrho)$ and two or three-setting linear steering $ \mathbb{S}_{2}(\varrho)$($ \mathbb{S}_{3}(\varrho)$) (a)under the local unital noise $\Gamma t$ with initial state $(c_{1},c_{2},c_{3})=(0.8,1,1)$ and (b) nonunital noise  $\Gamma t$ ($\Gamma=0.005\kappa$) with initial state $(c_{1},c_{2},c_{3})=(1,1,0.8)$.}
\end{figure}

\section{Average Correlation versus quantum steering under noisy channels}

To further gain insights into average correlation versus quantum steering, in this section we would like to compare their performances under the influence of local unital and nonunital noisy channels. Usually, the evolved state of such a system under local noises can be described as a completely positive trace preserving map, $\varepsilon[\rho]=\sum_{ij}(E_{i}\otimes E_{j})\rho(E_{i}\otimes E_{j})^\dagger$ with the local Kraus operators satisfying$\sum_{k}E_{k}^{\dagger}E_{k}=\openone$. For simplicity, we restrict ourselves to the case in which the initial states are prepared in the maximally mixed marginal states given by Eq.(1) with the Bloch vectors $r=s=0$. In what follows, we are devoted to address how do environmental noises influence the average correlation versus quantum steering.

\subsection{Unital noise}

We first consider a type of local unital noisy channels\cite{Inf. Theory 47(2001)} which satisfy the unital condition $\varepsilon(\frac{1}{2}\openone)=\sum_{\mu}E_{\mu} (\frac{1}{2}\openone)E_{\mu}^\dagger=\frac{1}{2}\openone$. Such as bit-flip, bit-phase-flip, and phase-flip channels belong to this category of noises. The corresponding Kraus operators are denoted by $E_{0}=\sqrt{1-p/2}\openone$ and $E_{1}=\sqrt{p/2}\sigma_{i}$, where $\sigma_{i}$ are the three Pauli's matrices. More specifically, $\sigma_{i}, (i=1,2,3)$ are corresponding to bit-flip, bit-phase-flip, and phase-flip channels, respectively. $p=1-e^{-\Gamma t}$ represents decoherence probabilities. Assume that the subsystem is subjected to the same quantum channel, the evolved state of such system is determined by
\begin{subequations}
\begin{eqnarray}
c_{i}'&=&c_{i},  \\
c_{j}'&=&c_{j}(1-p)^2, \mbox{ for } i\neq j.
\end{eqnarray}
\end{subequations}
Here $i=1,2,3$ represents the system suffered from bit-flip, bit-phase-flip, and phase-flip channels, respectively.
Using Eqs.(8)and (9), we immediately obtain the degree of steering under unital noisy channels
\begin{equation}
s_{2}(\varepsilon[\rho])=\sqrt{\sum_{i}^{i=3} |c_{i}^{'}|^2-\gamma^2}
\end{equation}
\begin{equation}
s_{3}(\varepsilon[\rho])=\sqrt{\sum_{i}^{i=3} |c_{i}^{'}|^2}
\end{equation}
where $\gamma=min\{|c_{i}^{'}|,|c_{j}^{'}|\}$.  At the same time, the average correlation under unital noisy channels is strongly depended on Eq.(6), where
$\alpha=max\{|c_{i}^{'}|,|c_{j}^{'}|\}$.
Obviously, both $\mathbb{S}_{n}(\xi[\varrho])\leq \mathbb{S}_{n}(\varrho)$ and $\Sigma(\xi[\varrho])\leq \Sigma(\varrho)$ are strictly satisfied due to the fact that $|c^{'}_{i}|\leq |c_{i}|$ $(i=1,2,3)$ always hold, this indicates
the unital noises never increase the amount of quantum steering and average correlations.

For an illustration, the behavior of average correlation and quantum steering against $\Gamma t$ under unital noisy channels is depicted in Fig.2(a), where we can clearly see that the amount of quantum steering and average correlations decreases monotonically under the above local unital channels. Moreover, the average correlations $\Sigma(\varepsilon[\varrho])$ is more robust against decoherence in comparison to $\mathbb{S}_{2}(\varepsilon[\varrho])$ and $\mathbb{S}_{3}(\varepsilon[\varrho]) $, and they obey the hierarchy of robustness
\begin{equation}
\mathbb{S}_{2}(\varepsilon[\varrho])\Rightarrow \mathbb{S}_{3}(\varepsilon[\varrho]) \Rightarrow\Sigma(\varepsilon[\varrho]).
\end{equation}

To support our statement for the hierarchy of robustness under arbitrary lossy channels, we result to the initial states $\vec{c}=(c_{1}=c_{2}=c_{3}=|c|)$, and let $t_{s_{2}}$, $t_{s_{3}}$ and $t_{\Sigma}$  be the sudden-death time or the shortest time where the corresponding $\mathbb{S}_{2}(\varepsilon[\varrho])$, $\mathbb{S}_{3}(\varepsilon[\varrho]) $ and $\Sigma(\varepsilon[\varrho])$ disappear, respectively,
\begin{subequations}
\begin{eqnarray}
t_{s_{3}}=-\frac{\ln[\frac{1-|c|^2}{2|c|^2}]}{4\Gamma},  \\
t_{s_{2}}=-\frac{\ln[\frac{1-|c|^2}{|c|^2}]}{4\Gamma},\\
t_{\Sigma}=-\frac{\ln A}{4\Gamma}.
\end{eqnarray}
\end{subequations}
Here $A$ satisfies the equation $Arcsinh[\sqrt{\frac{1-A}{A}}]=\frac{1-|c|}{|c|}\frac{\sqrt{1-A}}{A}$. As expected, for any fixed $c$, $t_{s_{2}}\leq t_{s_{3}}\leq t_{\Sigma}$ holds, this implies $\mathbb{S}_{2}(\varepsilon[\varrho])$ vanishes before $ \mathbb{S}_{3}(\varepsilon[\varrho])$, while $ \mathbb{S}_{3}(\varepsilon[\varrho])$ vanishes before $\Sigma(\varepsilon[\varrho])$. These analytical results give strong support to the hierarchy claimed by Eq.(28). However, the question naturally arises whether this phenomenon still occurs under the influence of nonunital
channels?

\subsection{Nonunital noise}

Generalized amplitude damping is usually regarded as a nonunital noisy channel\cite{Phys.Rev.Lett93(2004)} which is defined as $\varepsilon(\frac{1}{2}\openone)=\sum_{\mu}E_{\mu} (\frac{1}{2}\openone)E_{\mu}^\dagger=(1-p/2)|0\rangle\langle0|+p/2|1\rangle\langle1|\neq\frac{1}{2}\openone$.
The corresponding Kraus operators are denoted by $E_{0}=|0\rangle\langle0|+\sqrt{1-p}|1\rangle\langle1|$ and $E_{1}=\sqrt{p}|0\rangle\langle1|$. Here $p=1-e^{-\Gamma t}[\cos(\frac{Dt}{2})+\frac{\Gamma}{D}\sin(\frac{Dt}{2})]^2$ with $D=\sqrt{2\kappa\Gamma-\Gamma^2}$. In particular, for the case where $\Gamma<2\kappa$, the system is located in the strong-coupling regime, while for the case where $\Gamma>2\kappa$, the system is lied in the weak-coupling regime.  Similarly to unital cases, the evolved state under this nonunital channels is expressed as
\begin{subequations}
\begin{eqnarray}
c_{1,2}'&=&c_{1,2}(1-p),  \\
c_{3}'&=&c_{3}(1-p)^2+p^2
\end{eqnarray}
\end{subequations}
leading to the dynamics of average correlation and quantum steering are no longer monotonic.
Fig.2(b) demonstrates the dynamics of average correlation and quantum steering against $\Gamma t$ under nonunital noisy channels. It can be seen that
the decay and revival of average correlations and quantum steering occur. Most interesting, they follow a particular order: The order of decay satisfies $t_{s_{2}}\leq t_{s_{3}}\leq t_{\Sigma}$ which is the same hierarchy as in the case of unital channels, while the revival order follows in the reverse hierarchy $t_{s_{2}}\geq t_{s_{3}}\geq t_{\Sigma}$.

\section{Discussion and conclusion \label{sec:Conclusion}}

Before ending this paper, yet there are still some open theoretical problems for further consideration. One is the question that whether there
exists the similar relationship between the average correlation and the maximum violation of a general n-setting linear steering inequality for more general states. In addition, it remains an open question whether our results can be generalized to the bipartite states of higher dimensions.

In summary, we have established the connection between the average correlation and the violation of the three-setting linear steering inequality for two-qubit systems. We have analytically derived the extremal average correlation for a given amount of the three-setting linear steering and characterize the respective states. For clarity of our presentation, we have illustrated these results with examples from well-known classes of two-qubit states, including entangled pure and mixed states. The results show that average correlation is closely related to three-setting linear steering, like its relationship with Bell nonlocality quantified by the maximum violation of the Bell inequality. To further insight into their connection, we have also explored the dynamical behavior of these two quantifiers  under the influence of local unital and nonunital noisy channels. The results suggest that unital noisy channels never increase nonclassicality (quantified by average correlation) and nonlocality (quantified by the two or three-setting linear steering inequality), while nonunital noisy channels can induce revival of average correlations and quantum steering. Particularly, for a given class of states, there exists the aforementioned hierarchy
$\mathbb{S}_{2}(\rho)=\mathbb{N}_{2}(\rho)\Rightarrow \mathbb{S}_{3}(\rho) \Rightarrow\Sigma(\rho)$, as by the hierarchy it follows that the latter resource implies the former. In short, we hope the results in this paper would help us to better understand the relation between nonlocality and nonclassicality of the resource states.

\section{Acknowledgment}

This work is supported by the Natural Science Foundation of Hunan Province (Grant No.2021JJ30757). YN Guo is supported by the Program of Changsha Excellent Young Talents ( kq1905005, kq2009076, kq2106071, kq2206052 and kq2305029).

\end{document}